# The MACHO Project Dark Matter Search

D.P. Bennett[†,*], C. Alcock[*,†], R.A. Allsman[*], T.S. Axelrod[*],

K.H. Cook[*,†], K.C. Freeman[‡], K. Griest[†,∥], S.L. Marshall[†,♭],

B.A. Peterson[‡], M.R. Pratt[†,♭], P.J. Quinn[‡], A.W. Rodgers[‡],

C.W. Stubbs[†,♣], W. Sutherland[♠] (The MACHO Collaboration)

[*] *Lawrence Livermore National Laboratory, Livermore, CA 94550*

[†] *Center for Particle Astrophysics, University of California,*
*Berkeley, CA 94720*

[‡] *Mt. Stromlo and Siding Spring Observatories,*
*Australian National University, Weston, ACT 2611, Australia*

[∥] *Dept. of Physics, University of California, San Diego, CA 92039*

[♭] *Dept. of Physics, University of California, Santa Barbara, CA 93106*

[♣] *Dept. of Astronomy, University of Washington, Seattle, WA 98195*

[♠] *Dept. of Physics, University of Oxford, Oxford OX1 3RH, U.K.*

**Abstract.** We provide a status report on our search for dark matter in our Galaxy in the form of massive compact halo objects (or Machos), using gravitational microlensing of background stars. This search uses a very large format CCD camera on the dedicated 1.27m telescope at Mt. Stromlo, Australia, and has been taking data for almost 3 years. At present, we are in the midst of analyzing our second year data for 8 million stars in the Large Magellanic Cloud. We find more microlensing events than expected from known stellar populations suggesting that Machos are indeed present in the Galactic halo, but the observed microlensing rate toward the LMC is too small to allow for a halo dominated by sub-stellar Machos. Our observations of the Galactic bulge have also yielded substantially more microlensing events than anticipated including a number of exotic "deviant" microlensing events. The implications of these results are discussed.

## 1. Introduction

Perhaps the most important unsolved problem in astrophysics is the nature of the dark matter that dominates the Universe. There is evidence for dark matter at a range of scales, but the strongest evidence for dark matter is in the halos of spiral galaxies which dominate the galactic masses but contribute essentially no light. Many dark matter candidates, especially hypothetical elementary particles, have been proposed to account for the unseen mass, but the least speculative option is that the dark matter consists of stellar or planetary mass objects that are not burning nuclear fuel. Examples include jupiters, brown dwarfs, white dwarfs, and black hole remnants of collapsed stars, and they are collectively known as MACHOs for MAssive Compact Halo Objects.

An ingenious idea for detecting Machos was suggested by Paczyński in 1986. He suggested that Machos in the halo of our Galaxy might be detected by means of gravitational microlensing, and this possibility was confirmed in a rather spectacular fashion in September of 1993 when the EROS, MACHO, and OGLE collaborations each discovered their first candidate microlensing events in the course of less than a month (Alcock, *et al.*, 1993, Aubourg, *et al.*, 1993, and Udalski, *et al.*, 1993). Since then, the total number of microlensing events discovered has grown to about 80, and it has become clear that the Milky Way Halo cannot be dominated by Machos of sub-stellar mass. However, the microlensing optical depth toward both the Large Magellanic Cloud and the Galactic bulge seems to be larger than predicted on the basis of known stellar populations.

The basic physics of microlensing is quite simple. If a compact object passes very close to the line of sight to a background star, the light will be deflected to produce two images of the star. In the case of perfect alignment, the star will appear as an 'Einstein ring' with a radius of $r_E = \sqrt{4GMLx(1-x)/c^2}$, where $M$ is the lens mass, $L$ is the observer-source distance and $x$ is the ratio of the observer-lens and observer-source distances. In a typical situation of imperfect alignment, the image appears as two arcs. For lenses inside our Galaxy, the angular separation is $\lesssim 0.001$ arcsecond which is well below even the Space Telescope resolution, hence the term 'microlensing'.

However, in the point source approximation, the lensing produces a net amplification of the source by a factor

$$A = \frac{u^2 + 2}{u\sqrt{u^2 + 4}}$$

where $u = b/r_E$ and $b$ is the distance of the lens from the direct line of sight. The amplification is approximately $u^{-1}$ for $u \lesssim 0.5$, and $1 + 2u^{-4}$ for $u \gg 1$, hence the amplification can be very large but falls rapidly for $u \gtrsim 1$. Since objects in the Galaxy are in relative motion, this amplification will be time-dependent; for a typical lens transverse velocity of 200 km/s, the duration is $\hat{t} \equiv 2r_E/v_\perp \approx 140\sqrt{M/M_\odot}$ days. This is a convenient timescale for astronomical observations, and thus by sampling on a range of timescales the microlensing searches may be sensitive to a wide mass range from small planets of $\sim 10^{-6} M_\odot$ to black holes of $\sim 100 M_\odot$, covering most of the plausible candidates.

## 2. Optical Depth

The 'optical depth' $\tau$ for gravitational microlensing is defined as the probability that a given star is lensed with $u < 1$ or $A > 1.34$ at any given time, and is

$$\tau = \pi \int_0^L \frac{\rho(l)}{M} r_E^2(l)\, dl$$

where $l$ is the distance along the line-of-sight and $\rho$ is the dark matter density. Since $r_E \propto \sqrt{M}$, while for a given $\rho$ the number density of lenses $\propto M^{-1}$, the optical depth is independent of the individual MACHO masses. Using the virial theorem, it is found that $\tau \sim (v/c)^2$, where $v$ is the rotation speed of the Galaxy. More detailed calculations (Griest, 1991) give an optical depth for lensing by halo dark matter of stars in the Large Magellanic Cloud (LMC) of $\tau_{\text{LMC}} \approx 5 \times 10^{-7}$. Similar calculations led to the prediction of $\tau_{\text{bulge}} \approx 10^{-6}$ for lensing of bulge stars by low mass stars in the bulge and disk.

These very low values are the main difficulty of the experiment; only $\sim 1$ star in a million will be amplified by $A > 1.34$ at any given time, while the fraction of variable stars is much higher, $\sim 3 \times 10^{-3}$.

## 3. Microlensing Signatures

Fortunately, microlensing has many strong signatures which can discriminate it from stellar variability: 1) Since the optical depth is so low, only one event should be seen in any given star. 2) The deflection of light is wavelength-independent, so the star should not change color during the amplification, although in cases of severe blending of stellar images, very good seeing images may be required to determine the unlensed color of the lensed star. 3) The accelerations of galactic objects are generally negligible on timescales of these events, hence the events should be symmetrical in time and have the shape $(A(u))$ given above with with $u(t) = \sqrt{u_{min}^2 + (v_\perp(t - t_{max})/r_E)^2}$. Examples of such light-curves are shown in Figures 1 and 5. All these characteristics are distinct from known types of intrinsic variable stars; most variable stars are periodic or semi-regular, and do not remain constant for long durations. They usually change temperature and hence color as they vary, and they usually have asymmetrical lightcurves with a rapid rise and slower fall.

In addition to these individual criteria, if many candidate microlensing events are detected, there are further statistical tests that can be applied: 4) The events should occur equally in stars of different colors and luminosities, 5) the distribution of impact parameter $u_{min}$ should be uniform from 0 to the experimental cutoff $u(A_{min})$, and 6) the event timescales and peak amplifications should be uncorrelated. (In practice these distributions will be modulated by the detection efficiencies, which can be computed from simulations).

## 4. Microlensing Searches

Due to the low optical depth, a very large number of stars must be monitored over a long period to achieve a significant detection rate. The optimal targets when searching for Machos in the galactic halo are the Large and Small Magellanic Clouds, the largest of the Milky Way's satellite galaxies, since they have a high surface density of stars and are distant enough at 50 and 60 kpc to provide a good path length through the dark halo.

Another target for microlensing searches is the Galactic bulge where a higher microlensing optical depth is predicted due to lensing by faint stars in the Galactic disk (Griest, *et al.*, 1991, Paczyński, 1991). Naively, one might expect that microlensing toward the bulge would not be of great relevance for the halo dark matter problem, but as we shall see, this is not the case.

There are currently four microlensing surveys that have reported results: the EROS, MACHO, OGLE, and DUO collaborations. The EROS collaboration operates two separate experiments which observe the LMC. They have searched for events lasting from days to months using photographic plates taken at the ESO Schmidt and for events lasting less than a day from a 40-cm telescope with a large format CCD array camera at ESO. The OGLE collaboration has observed the Galactic bulge with $\sim 75$ nights per year on the 1-m Swope telescope at Las Campanas, Chile. Both the EROS and OGLE teams are presently upgrading to dedicated 1-m class telescopes in Chile. The DUO collaboration has observed the bulge using the ESO Schmidt for a single observing season.

The MACHO collaboration has full-time use of the 1.27-m telescope at Mt.Stromlo Observatory, which images 0.5 square degrees of sky simultaneously in 2 color bands. The two foci are equipped with large CCD cameras, each containing 4 Loral CCD chips of $2048 \times 2048$ pixels. The observing time is divided between the LMC and bulge, with a small amount of time being spent on the SMC.

## 5. Microlensing Results Towards the LMC

The analysis of our first year LMC data has recently been completed (Alcock, *et al.*, 1995a, 1995b), and we are presently in the midst of analyzing our second year LMC data. Here, we will summarize the results of our first year LMC data and briefly discuss some of the results of the year-2 analysis.

The first year data consist of 22 half square degrees fields located in or near the LMC bar that were observed between 140 and 350 times between August, 1992 and September, 1993. The analysis of the 8.6 million two color light curves in this data set is described in some detail in Alcock *et al.*(1995b). Our analysis yielded 4 microlensing-like lightcurves, but two of these represent the same star in a field overlap region. Thus, we have detected the 3 candidate microlensing events whose lightcurves are shown in Figure 1.

Quantitative interpretation of these results depends upon the event detection efficiency of our experiment. Inefficiencies arise because of 1) incomplete sampling of the light curves (mostly due to weather) and 2) because of the unresolved blending of stars in an image, only one of which will be lensed. In the case of blending, the efficiency per target is reduced because of distortion of the light curve and dilution of the amplification; at the same time the number of real target objects is increased, which increases the effective efficiency. These two effects partially offset each other. Our method for calculating our detection efficiency involves adding artificial stars into real images in order to take both sampling and blending effects into account (Alcock *et al.*1995a, 1995b).

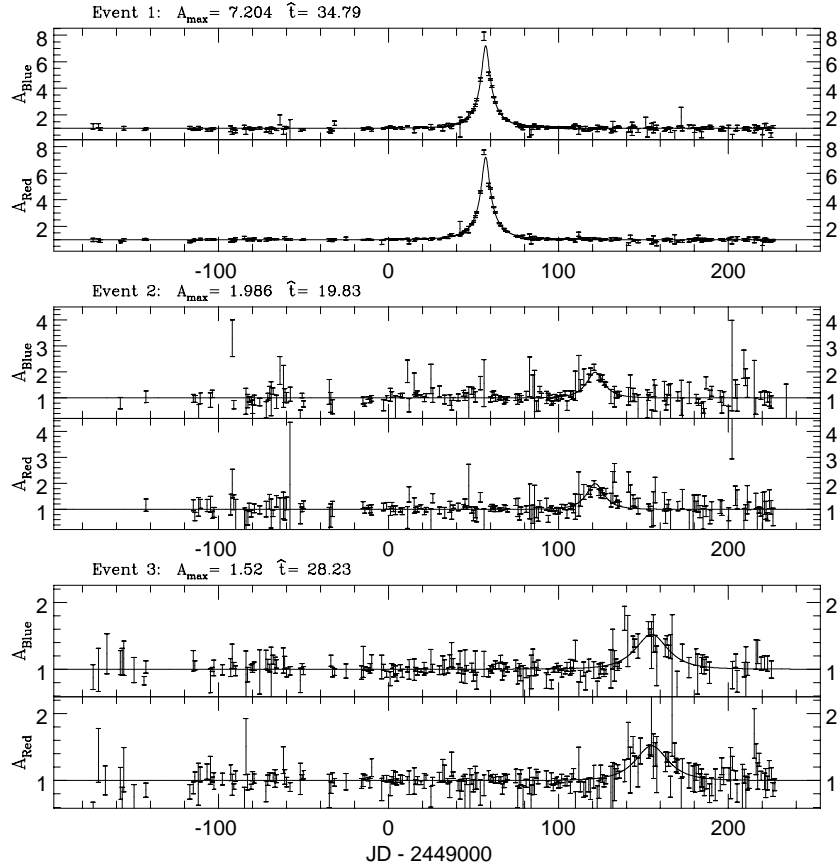

Figure 1. Light curves for the 3 candidate events in the LMC. Flux is in linear units, relative to the unamplified flux. The smooth curves are the microlensing fit to both colours simultaneously.

In order to illustrate the implications of this result we have compared the number and timescales of the observed events with predictions of the commonly used spherical model of the density of dark matter in the Galaxy's halo, $\rho_H(r) = \rho_0(r_0^2 + a^2)/(r^2 + a^2)$, where $r$ is galactocentric radius, $r_0 = 8.5\,\text{kpc}$ is the galactocentric radius of the Sun, $a = 5\,\text{kpc}$ is the halo core radius and $\rho_0 = 0.0079 \text{M}_\odot \text{pc}^{-3}$ is the local dark matter density. This model gives a Galactic halo mass of $4.1 \times 10^{11} \text{M}_\odot$ within $50\,\text{kpc}$ of the Galactic center. After incorporating the detection efficiency, the predicted number of detected events $N_{exp}(m)$ for a dark halo entirely composed of compact objects with a unique mass $m$ is shown in Figure 3. There are two ways to interpret our results. First, we can conservatively interpret all three events as background and use Poisson statistics to rule out at 95% CL all halo models which predict in excess of 7.7 events. Defining $\psi(m)dm$ as the mass fraction of halo Machos with masses between $m$ and $m + dm$, Figure 2 shows that models with $\psi(m)$ equal to a delta function are ruled out for masses between $m_{low} = 8 \times 10^{-5} \text{M}_\odot$ and $m_{high} = 0.3 M_\odot$. In fact, a model with *any* mass distribution restricted to this excluded mass range is also ruled out, since $N_{exp}(m) > 7.7$ implies $N_{exp} = \int_{m_{low}}^{m_{high}} \psi(m) N_{exp}(m) dm > [7.7 \int_{m_{low}}^{m_{high}} \psi(m) dm = 7.7]$. So lens objects in the above mass range cannot contribute 100% of the model halo at 95% CL. Taking our result in conjunction with the null results from the EROS collaboration's CCD experiment (Aubourg *et al.*, 1995), these two microlensing experiments have placed strong constraints on a dark halo of lensing objects of the form we have assumed for masses between $5 \times 10^{-8} M_\odot$ and $0.3 M_\odot$.

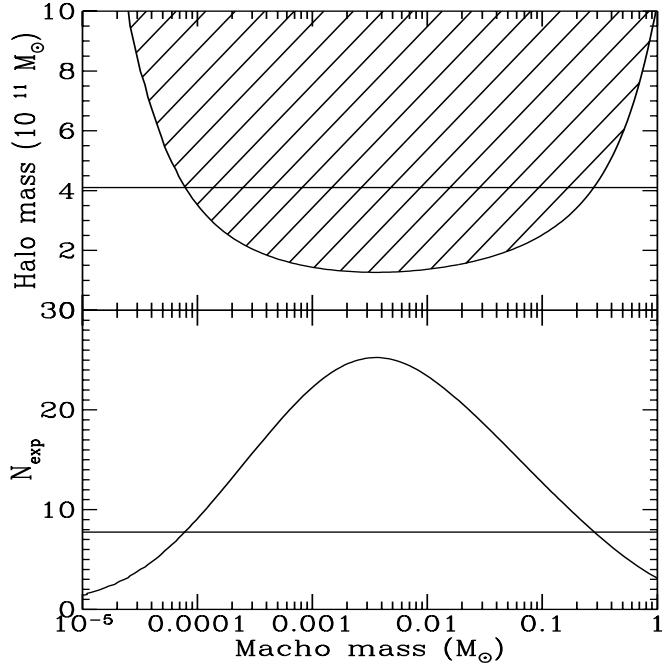

Figure 2. The lower panel shows the number of expected events predicted from the standard model halo with a delta function mass distribution. Given three observed events, points above the line drawn at $N_{exp} = 7.7$ are excluded at the 95% CL. The upper panel shows the 95% CL limit on the halo mass in MACHOs within 50 kpc of the galactic center for the model. Points above the curve are excluded at 95% CL while the line at $4.1 \times 10^{11} M_\odot$ shows the total mass in this model within 50 kpc.

We have explored a range of different halo density profiles (Alcock *et al.*, 1995b), and we find that while the constraint on the halo mass *fraction* in MACHOs is quite model-dependent, our constraints on the *total* mass of MACHOs interior to 50 kpc are relatively independent of the assumed model of the galactic halo. It is still possible to construct models in which the halo is made entirely of MACHOs, but these models require a small galactic mass inside 50 kpc and are only marginally consistent with rotation curve and satellite galaxy constraints (Kohchanek, 1995).

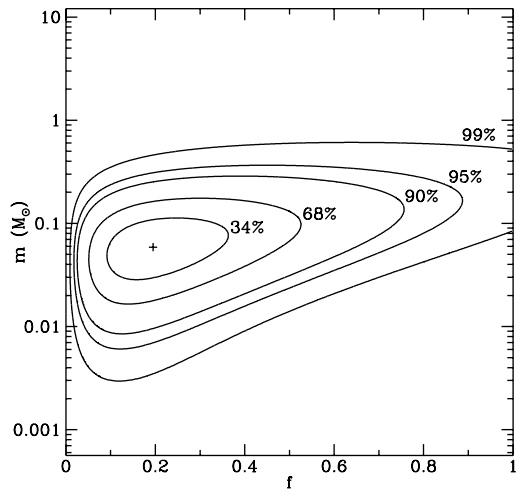

Figure 3. Likelihood contours for the model halo with a delta function mass distribution. The numbers labeling the contour levels indicate the total probability of the enclosed parameter region according to a Bayesian analysis assuming a prior distribution uniform in $m$ and $f$. The position of the most likely MACHO mass $m$ and MACHO halo fraction $f$ is marked with a +.

In setting the limits above we need not assume that our three events are due to microlensing by halo objects; however, if we add this assumption we can use the observed durations of the events in a maximum likelihood analysis to find the most likely MACHO fraction $f$, and mass $m$ (again with a delta function mass distribution). In this model the rest of the halo is presumed to consist of objects with masses outside our range of sensitivity. Figure 3 shows the likelihood contours with most likely values of $f_{2D} = 0.17$ and $m_{2D} = 0.04 M_\odot$. The one-dimensional confidence interval for $f$ is $f_{ML} = 0.19^{+0.16}_{-0.10}$, where the errors are 68% CL.

The model of equation 2 predicts an optical depth of $\tau_{model} = 4.7 \times 10^{-7}$. Using $f_{ML}$ we can now state the model best fit optical depth of $8.8^{+7}_{-5} \times 10^{-8}$ and observed MACHO mass within 50 kpc of $7.6^{+6}_{-4} \times 10^{10} M_\odot$. We can also estimate the optical depth directly without model assumptions which yields: $\tau_{est} = 8.0 \times 10^{-8}$, in good agreement with the maximum likelihood estimate.

It is possible that the observed events, if microlensing, might be due to objects in the LMC (Wu 1994, Sahu 1994) or a non–halo Galactic population (Gould et al., 1994). We have estimated the event rates due to known stars in Galactic and LMC populations and find that they should contribute on average $\sim \frac{1}{2}$ an event to our sample of 3. For three detected events the 90 % CL lower bound on the underlying rate is 1.1 events. Thus, it seems likely that a new Galactic population, probably in the halo, is responsible for the detected microlensing events (unless two of the three events are not due to microlensing).

It is expected that the analysis of the MACHO year-2 LMC data will shed some light on this situation, but this analysis is not complete enough to draw any definitive conclusions. It has, however, yielded some intriguing results. The 2+ year baseline will yield increased sensitivity due to the increased exposure time, but it will also allow the detection of longer time scale events that we were not previously sensitive to. Indeed, the year-2 analysis has yielded a few events with longer time scales than those seen in year-1. The most spectacular example of a recently discovered longer timescale event is shown in Figure 4. This event appears to have a rather short timescale because its amplification is very high– the point at $A \gtrsim 40$ is the highest amplification ever observed in a lensing event, but the Einstein diameter crossing time $\hat{t} = 110$ is three times longer than any of the events shown in Figure 1. The implications of these newly discovered longer timescale events are still quite uncertain, but it seems likely that our improved sensitivity to longer timescale events will result in a microlensing optical depth larger than the $\tau = 8 \times 10^{-8}$ value from the year-1 analysis.

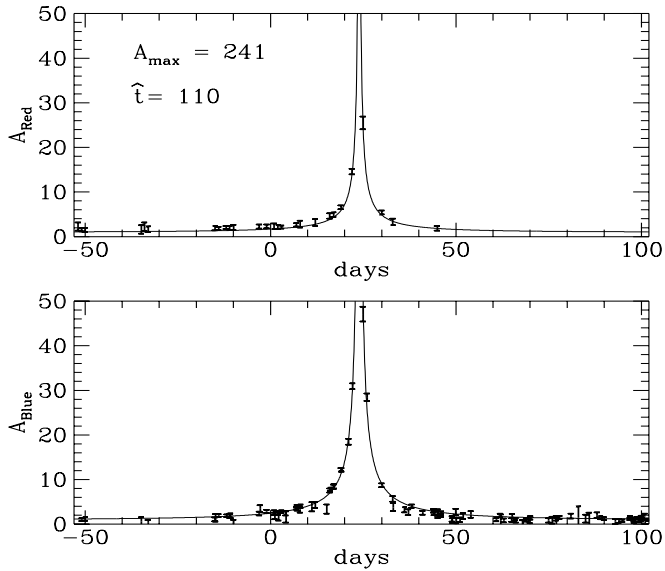

Figure 4. Lightcurve of a very high amplification microlensing event discovered in the MACHO year-2 LMC analysis. The fit curve shown includes the effects of blending of the lensed star with a very red unlensed source.

## 6. Microlensing Toward the Galactic Bulge

Within a week after the EROS and MACHO teams announced the first microlensing events detected toward the LMC, the OGLE collaboration had discovered the first microlensing event ever seen toward the Galactic bulge (Udalski, et al., 1993). Subsequent analyses by the OGLE group revealed about 10 additional events, and revealed a very high optical depth towards Baade's window (Udalski, et al., 1994). These results are fully consistent with the MACHO results that will be discussed in this section.

We have recently analyzed the 2-color light curve data for about 12 million stars in 24 fields toward the Galactic bulge. Our microlensing search of these data revealed 45 candidate microlensing events concentrated towards the Galactic plane (Bennett *et al.*, 1995) with timescales ranging from $\hat{t} = 9$ to $\hat{t} = 200$ days. This sample includes the exotic events discussed in the next section.

The interpretation of the bulge events is complicated by the fact that an uncertain fraction of the stars we see towards the bulge are actually foreground members of the galactic disk. The microlensing optical depth is quite a bit smaller for these stars, so they make it more difficult to measure the optical depth to the stars which actually reside in the bulge. We have chosen to avoid this problem by concentrating on a subset of 13 lensing events in which the source star is a "red clump" giant. These stars have the virtue that they are very likely to be located in the Bulge, and they are bright enough so that blending does not make an important contribution to the microlensing detection efficiency for clump giant source stars. The 13 lensing events detected among the 1.3 million clump giant stars yield a microlensing optical depth of $\tau_{\text{bulge}} = 3.9^{+1.8}_{-1.2} \times 10^{-6}$ averaged over an area of $\sim 12$ square degrees centered at Galactic coordinates $\ell = 2.55°$ and $b = -3.64°$. This is quite similar to the value reported by the OGLE collaboration, but it is much higher than the model predictions of $\tau_{\text{bulge}} \sim 10^{-6}$ (Griest *et al.*, 1991, Paczyński, 1991). This suggests that the mass in low mass stars along the line of sight toward the bulge is considerably higher than previously thought. This could come about if the mass of the disk was larger than in standard galactic models (Alcock, et.al. 1995c) or if the Galactic bulge is really a bar pointing nearly along the line of sight (Zhao, Spergel, and Rich, 1995, Paczyński, *et al.*, 1995). Both of these possibilities improve the situation, but the predicted optical depths in these models tend to be more than one $\sigma$ below the measured value.

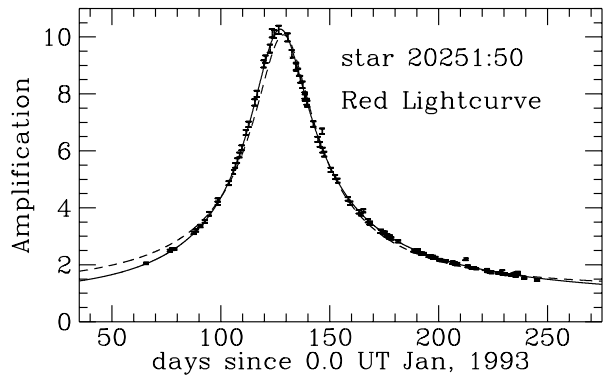

Figure 5. The MACHO-Red band light curve for the longest event yet detected by the MACHO project. The solid curve is the fit light curve which includes the effect of the earth's motion, while the dashed curve is the best fit ignoring the motion of the earth.

The nature of the bulge lensing events will have important implications for the dark matter problem since at least one of the proposed explanations for the high optical depth (the heavy disk) requires that almost all of the mass of the inner galaxy be in disk and not in the halo. This means that the halo would have to have a very large core radius which would be difficult to accommodate in cold dark matter models (Moore, 1994). In the 1995 bulge observing season we have extended our coverage in Galactic longitude in an attempt to separate the microlensing contributions of the disk and the Bulge.

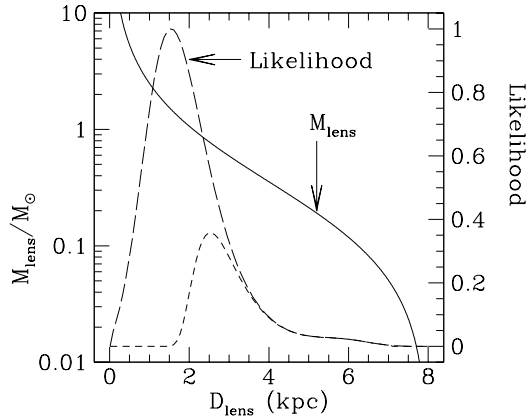

Figure 6. The solid curve shows lens mass is plotted as a function of the lens distance using the $v_t$ value determined by the light curve fit shown in Fig. 6. The long-dashed curve indicates the likelihood function for the lens distance incorporating the projected velocity and the Galactic model, while the short-dashed curve also includes the upper limit on the brightness of a main-sequence lens.

## 7. Deviant Microlensing

Another way to try and separate the disk and bulge lensing populations is to study events which allow us to learn additional details about the properties of the lenses. In most lensing events, the only information about the lens mass, distance and velocity that is available is all folded into a single parameter, the timescale $\hat{t}$. However, if one can measure details of the deviation of a microlensing light curve from the usual simple symmetric form, one can get an additional handle on the event.

A good example of this is shown in Figure 5 which is the light curve of the longest event that we have detected. This event displays a noticeable asymmetry which can be fit very well if we allow for the motion of the Earth around the Sun during the event. This is known as microlensing parallax (Refsdal, 1966; Gould 1992), and it allows us to determine additional parameters of the lensing event. The "parallax fit" reveals that the transverse velocity of the lens with respect to the source-observer line is $75 \pm 5$ km/s pointing in a direction which is 28 degrees from the direction of the Sun's motion about the Galactic center. This is exactly what one would expect for a lens residing in the disk.

The velocity determination allows us to obtain the mass distance relation that is shown in Figure 6. Further information about the lens can be obtained by considering the velocity distribution of the lensing objects which is fairly well known. Applying a likelihood analysis to the parameters of the parallax event and the velocity distributions of the disk and bulge gives the long-dashed curve shown in Figure 6. This likelihood function yields estimates of the distance to and mass of the lens: $D_{\rm lens} = 1.7^{+1.1}_{-0.7}$ kpc and $M = 1.3^{+1.3}_{-0.6} {\rm M}_\odot$. If we make the additional assumption that the lens is a main sequence star, then we can obtain an additional constraint on the properties of the lens shown by the short dashed curve in Figure 6 because no light can be detected from the lens. This last constraint actually rules out much of the region that was preferred by the velocity distribution consideration only. This suggests that either the lens is not a main sequence star (*i.e.* it's a white dwarf or neutron star), or it is located at $D_{\rm lens} = 2.8^{+1.1}_{-0.6}$ kpc with a mass of $M = 0.6^{+0.4}_{-0.2} {\rm M}_\odot$.

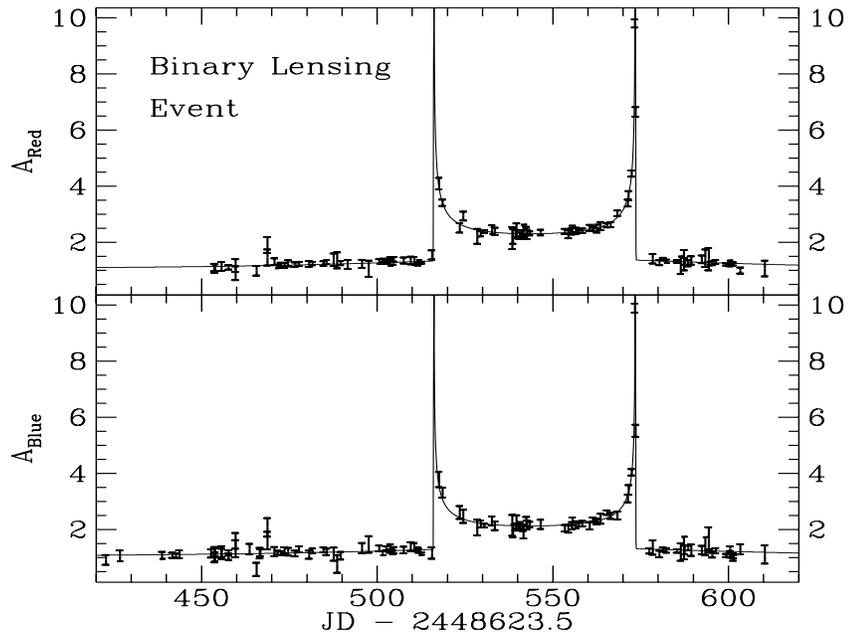

Figure 7. The MACHO lightcurve of a binary microlensing event first seen by the OGLE experiment.

Another, very spectacular deviant microlensing event in the MACHO year-1 bulge data is the binary microlensing event shown in Figure 7. This event occurred in the subset of our fields surveyed by the OGLE experiment and was first noticed by them (Udalski *et al.*, 1994b). Their data had a point at $A \sim 10$ on the first caustic crossing (*i.e.* the first light curve spike), but very poor coverage of the second caustic crossing, so they were rather cautious about interpreting this as a binary microlensing event. Our data, shown in Figure 7, clearly confirms that this is a binary microlensing event. The lenses in this event have nearly equal mass, but it is also possible to detect large deviations in microlensing event light curves due to planets as small as the Earth (Mao and Paczyński, 1991; Gould and Loeb, 1992). In fact, microlensing is probably the most cost effective technique for detecting extra-solar earth-mass planets.

## 8. Conclusions

The MACHO microlensing survey of the LMC has provided a stringent constraint on the total mass of sub-stellar objects that might reside in the Galactic halo, but it has also provided tantalizing evidence that Machos may contribute a significant fraction ($\sim 20\%$) of the total halo mass. The lensing rate seen towards the Galactic bulge is substantially higher than expected and this may have important implications for the dark matter composition as well. The large number of events seen towards the bulge have also resulted in a number of exotic microlensing events and have raised the possibility that microlensing might become a useful tool for studying the prevalence of extra-solar planets.

**Acknowledgments.** We are grateful to S. H. Rhie for the development of the binary microlensing event fitting code used for Figure 7. We also acknowledge the generous support of the DOE and NSF through contracts W7405-ENG-48 at LLNL and AST-8809616 at the CfPA.